\documentclass[12pt]{article}
\usepackage{amsfonts}
\usepackage{amssymb}
\usepackage{latexsym}
\usepackage{graphicx}
\usepackage[english]{babel}
\begin{document}
\input epsf

\def\p{\partial}
\def\h{{1\over 2}}
\def\be{\begin{equation}}
\def\bea{\begin{eqnarray}}
\def\ee{\end{equation}}
\def\eea{\end{eqnarray}}
\def\d{\partial}
\def\la{\lambda}
\def\eps{\epsilon}
\def\bb{\bigskip}
\def\mm{\medskip}
\newcommand{\dm}{\begin{displaymath}}
\newcommand{\edm}{\end{displaymath}}
\renewcommand{\b}{\tilde{B}}
\newcommand{\gm}{\Gamma}
\newcommand{\ac}[2]{\ensuremath{\{ #1, #2 \}}}
\renewcommand{\ell}{l}
\newcommand{\z}{\ell}
\newcommand{\newsection}[1]{\section{#1} \setcounter{equation}{0}}
\def\bb{$\bullet$}
\def\Qbar{{\bar Q}_1}
\def\QPbar{{\bar Q}_p}

\def\q{\quad}

\def\bn{B_\circ}

\let\a=\alpha \let\b=\beta \let\g=\gamma \let\d=\delta \let\e=\epsilon
\let\c=\chi \let\th=\theta  \let\k=\kappa
\let\l=\lambda \let\m=\mu \let\n=\nu \let\x=\xi \let\r=\rho
\let\s=\sigma \let\t=\tau
\let\vp=\varphi \let\vep=\varepsilon
\let\w=\omega      \let\G=\Gamma \let\D=\Delta \let\Th=\Theta
                     \let\P=\Pi \let\S=\Sigma

\def\h{{1\over 2}}
\def\t{\tilde}
\def\r{\rightarrow}
\def\nn{\nonumber\\}
\let\bm=\bibitem
\def\Kt{{\tilde K}}
\def\b{\bigskip}

\let\p=\partial

\begin{flushright}
\end{flushright}
\vspace{20mm}
\begin{center}
{\LARGE  Falling into a black hole}
\\
\vspace{18mm}
{\bf  Samir D. Mathur\footnote{mathur@mps.ohio-state.edu} }\\

\vspace{8mm}
Department of Physics,\\ The Ohio State University,\\ Columbus,
OH 43210, USA\\
\vspace{4mm}
\end{center}
\vspace{10mm}
\thispagestyle{empty}
\begin{abstract}

String theory tells us that quantum gravity has a dual description as a field theory (without gravity). We use the field theory dual to ask what happens to an object as it falls into the simplest black hole: the 2-charge extremal hole. In the field theory description the wavefunction of a particle is spread over a large number of `loops', and the particle has a well-defined position in space only if it has the same `position' on each loop. For the infalling particle we find one definition of `same position' on each loop, but there is a different definition for outgoing particles and no canonical definition in general in the horizon region. Thus the meaning of `position' becomes ill-defined inside the horizon. 

\end{abstract}
\newpage
\setcounter{page}{1}

When an object falls through a horizon, where does it go? This question leads to vexing paradoxes. But perhaps the situation is similar to the early days of quantum mechanics, when classical intuition gave questions with no resolution.
In this essay we take a simple explicit model of a black hole in string theory. We note that  gravitational systems have a {\it dual} description in terms of a  {\it field theory} \cite{maldacena}. Using the field theory dual we find that the notion of spacetime itself breaks down behind the horizon.

\b\b

{\large{\bf {The system}}}

\b\b

Compactify 10-d spacetime as
$$M_{9,1}~\r~M_{4,1}\times {\cal M}_4\times S^1$$
where the 4-manifold ${\cal M}_4$ is $T^4$ or $K3$. 
Wrap a string $n_1$ times around $S^1$. Wrap $n_5$ units of its electromagnetic dual, the fivebrane, on ${\cal M}_4\times S^1$. The union of these charges creates an {\it effective string} with winding
$$N=n_1n_5$$
around $S^1$. This effective string describes the field theory system \cite{maldasuss,lm4}. Its total length can be broken into loops in different ways (fig.1), with each loop having a spin (indicated by an arrow). For ${\cal M}_4=K3$ there are
$${\cal N}=e^{4\pi \sqrt{N}}\equiv e^{S_{micro}}$$
states arising from the allowed partitions \cite{sen}. 

On the other hand consider a spherically symmetric solution of Einstein's equations with these string and fivebrane charges. We get an extremal black hole whose horizon gives a Bekenstein-Wald entropy \cite{dabholkar}
$$S_{bek}~=~S_{micro}~.$$

\b\b

{\large{\bf The microstates}}

\b\b

The actual microstates of the system are however {\it not} spherically symmetric \cite{lm4}. We depict field theory states in fig.1. In fig.1(a) we have all loops of the same length and with the same spin. In the dual gravity description this brane state creates the geometry fig.2(a). We have flat space at infinity, then a deep throat. The throat is `capped off' smoothly, but spherical symmetry is broken to axially symmetry in a direction defined by the spins on the loops. 

\b\b

\begin{figure}[htbp] 
   \centering
   \includegraphics[width=6in]{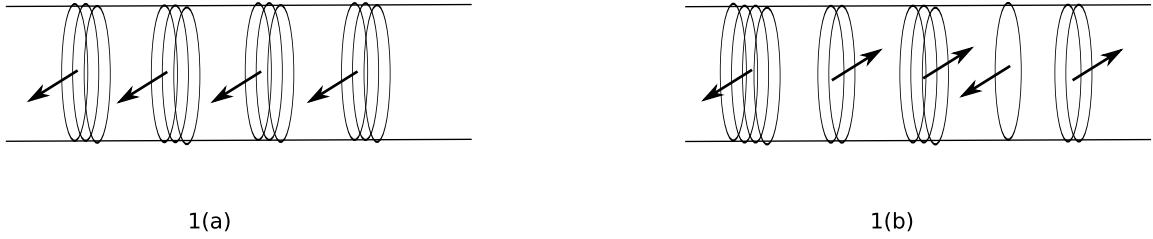} 
   \caption{Field theory states}
\end{figure}

\b\b

If we have loops of a few different kinds then the caps are more complicated (fig.2(b,c)). In the {\it generic} microstate  (fig.1(b)) the number of loops $N(m)$ having winding $m$ is given by a thermal distribution \cite{sen}
$$N(m)\sim {1\over e^{m\over \sqrt{N}}-1}$$
Thus the mean winding and the `spread in winding' are comparable
$$\bar m~\sim~\Delta m~\sim~ \sqrt{N}$$
and for loops of typical length $m\sim \bar m$
\be
N(m)~\sim ~1
\label{two}
\ee
Let $q$ be the number of loops in the state. For the generic state
$$q~\sim~ \sqrt{N} ~\gg~1$$

\begin{figure}[htbp] 
   \centering
   \includegraphics[width=6in]{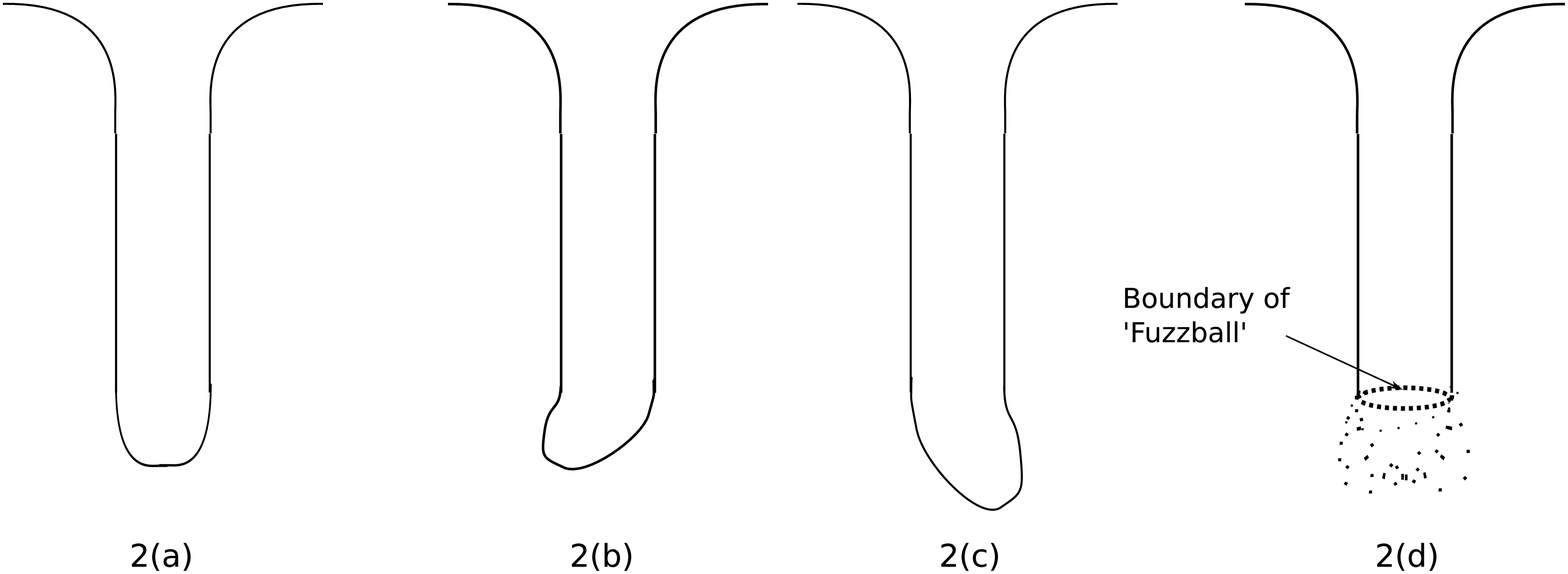} 
   \caption{Gravity solutions for different microstates}
\end{figure}

The caps are now very complicated, and because of (\ref{two}), also non-classical, i.e. very `quantum fluctuating' (fig.2(d)).  If we evaluate the area $A$ of the throat at the  location where  typical microstates start to differ from each other  (the dashed circle in fig.2(d)), then we find \cite{lm5}
$${A\over G}~\sim ~S_{micro}~=~S_{bek}$$
so the generic state is a horizon sized `quantum fuzzball'.

\b\b

{\large{\bf Infall in the classical geometry}}

\b\b

Let us now see what happens to a quantum that falls into the classical type geometry fig.2(a). In fig.3 we track the progress of the quantum using the field theory dual  fig.1(a).

\b\b

\begin{figure}[htbp] 
   \centering
   \includegraphics[width=6in]{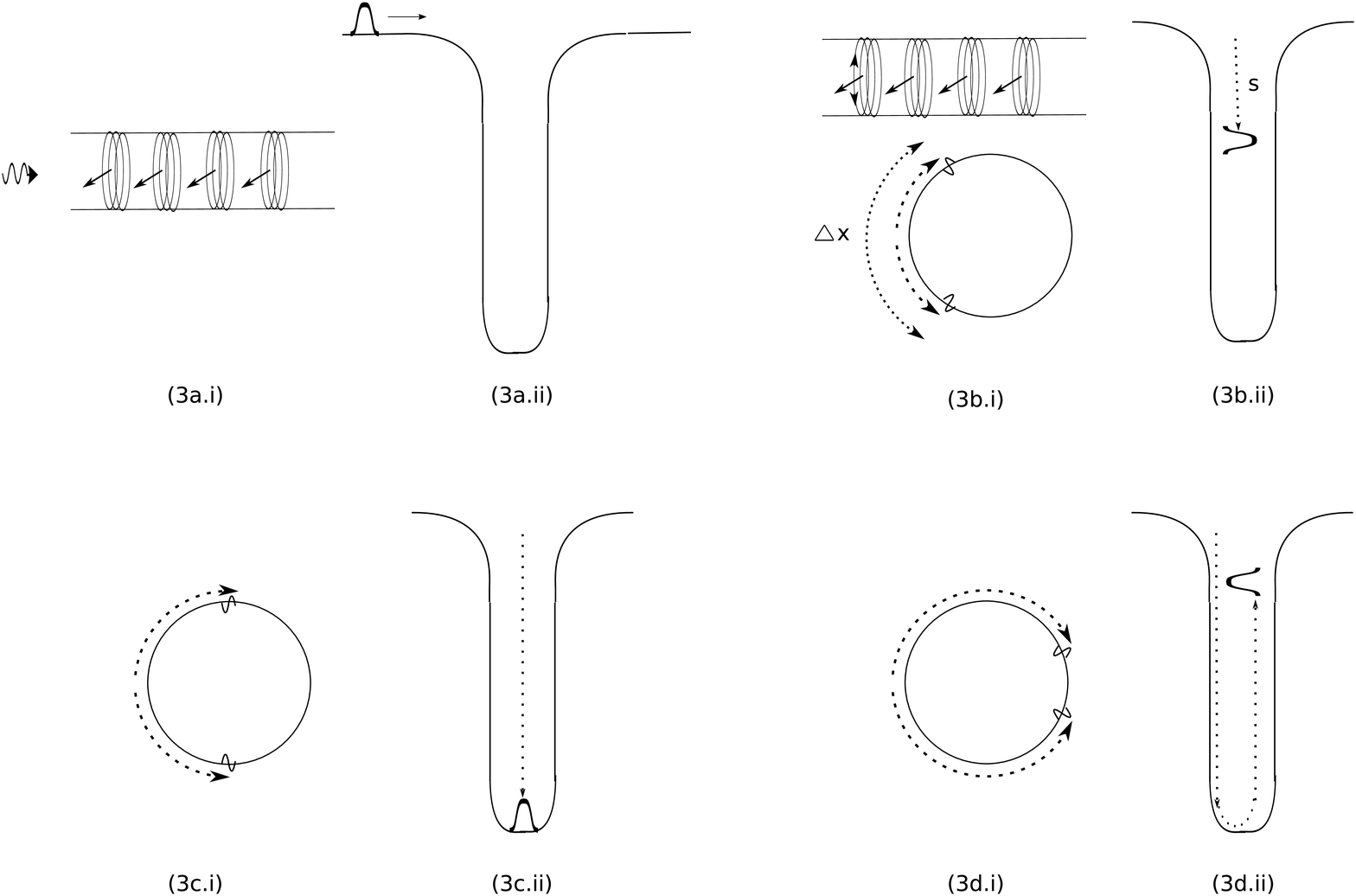} 
   
   \b
   
   \caption{Evolution in the classical state 1(a), 2(a).}
\end{figure}

\b\b

{\bf Fig.3(a):}\quad  Start with a quantum away from the brane state; in the gravity picture this quantum will be outside the throat. 

\b
{\bf Fig.3(b):}\quad In the field theory picture the branes absorb the quantum with a probability $P_{micro}$. One of the loops gets a pair of excitations that travel in opposite directions at the speed of light. Since any loop can be excited, the state of the branes is
\be
\Psi={1\over \sqrt{q}}~[({\rm loop ~'1'~excited})~+~({\rm loop ~'2'~excited})~+~\dots ~+~({\rm loop ~'q'~excited})]
\label{one}
\ee
We draw the excited loop `opened up to a circle'. After  time $\Delta t$ the excitations have separated by a distance 
$$\Delta x= 2 \Delta t$$
 
In the gravity dual   the quantum falls into the throat with probability $P_{gravity}=P_{micro}$, and moves down at the speed of light. In time $\Delta t$ it has gone down a distance
\be
s=\Delta t={\Delta x\over 2}
\label{three}
\ee

\b

{\bf Fig.3(c):}\quad The excitations on the loop reach diametrically opposite points. In the gravity dual the quantum reaches the `cap' and bounces back.

\b

{\bf Fig.3(d):}\quad The excitations on the loop approach each other again, and when they meet there is a probability $P_{micro}$ that they collide and exit the brane state (with probability $1-P_{micro}$ they continue around the loop). In the gravity dual the quantum moves up the throat, and with probability $P_{gravity}=P_{micro}$ exits to infinity (with probability $1-P_{gravity}=1-P_{micro}$  it reflects back into the throat).

\b\b

To summarize, the field theory description of spacetime involves a  large number of loops $q\sim\sqrt{N}\gg 1$, and the wavefunction of the particle is spread over all the loops. A second quantum on the spacetime will  also generate an excitation like (\ref{one}) but the probability that its excitation lies on the {\it same} loop is $\sim 1/q<<1$. Interestingly,  $\sim 1/q$ is the order of the gravitational interaction between the quanta. Also, the backreaction of the quantum of the geometry is $\sim 1/q$, so $q\gg1$ allows the quantum to be a `test particle' \cite{lm4}.  
 
 \b\b

{\large{\bf Infall in the generic microstate}}

\b\b

A black hole is described by  {\it generic} microstates (fig.1(b)) instead of  special ones like fig.1(a). 
So to study infall into the black hole fuzzball fig.2(d) we consider absorption into its dual fig.1(b). 

\b

The analogue of (\ref{one}) is now
\be
\Psi_{in}=[\alpha_1({\rm loop ~'1'~excited})~+~\alpha_2({\rm loop ~'2'~excited})~+~\dots ~+~\alpha_q({\rm loop ~'q'~excited})]
\label{onep}
\ee
where the amplitude for excitation of a loop with winding $m$ is found to be proportional to $m$. 
In fig.4 we sketch three loops  of different lengths.  For small times $\Delta t$ the evolution is similar to that  in fig.3: the excitations on any loop separate  by  $\Delta x=2\Delta t$ and the quantum in the dual geometry falls a distance $s=\Delta t$.
\b

But at larger times $\Delta t$ the excitations will have the locations sketched in fig.4.   On the largest loop, the excitations are not half-way around, and in the gravity dual would imply a quantum moving down towards the `cap'. In the middle sized loop, they are half-way around and in the smallest loop, the excitations have travelled past the half-way mark and in the gravity dual would correspond to a quantum moving back up the throat. 

\b

So {\it where} is the particle?

\b\b

\begin{figure}[htbp] 
   \centering
   \includegraphics[width=4.5in]{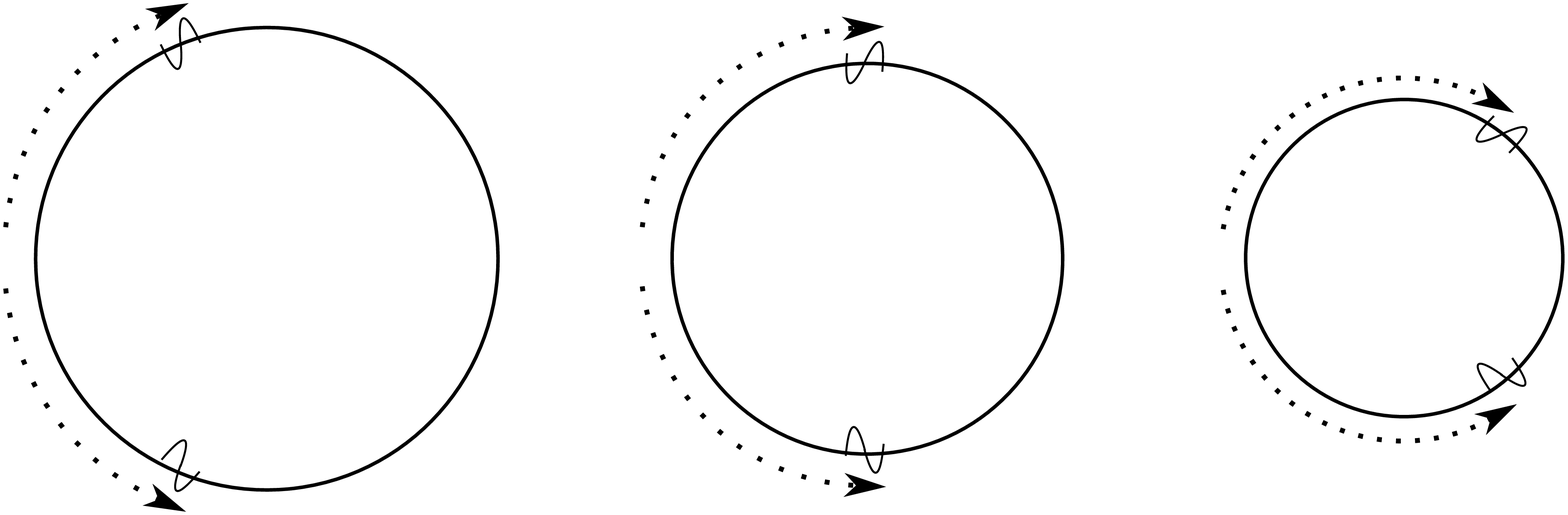} 
   \caption{The state $\Psi_{in}$; the excitations have moved the same distance on each loop}
\end{figure}

\b\b

{\large{\bf Different definitions of position}}

\b\b

Let us contrast the state $\Psi_{in}$ in fig.4 with a state $\Psi_{out}$ which would describe an {\it outgoing} particle with well-defined spacetime position (fig.5(a)). 
{\it On each loop the excitations are  the same distance away from recollision}. Since the recollision times are synchronized, the emitted wave from each loop is in phase, and the probability for emission of the quantum from the brane state works out to $P_{micro}$. On the gravity side the particle exits the throat with the same probability (fig.5(b)).  

\begin{figure}[htbp] 
   \centering
   \includegraphics[width=6in]{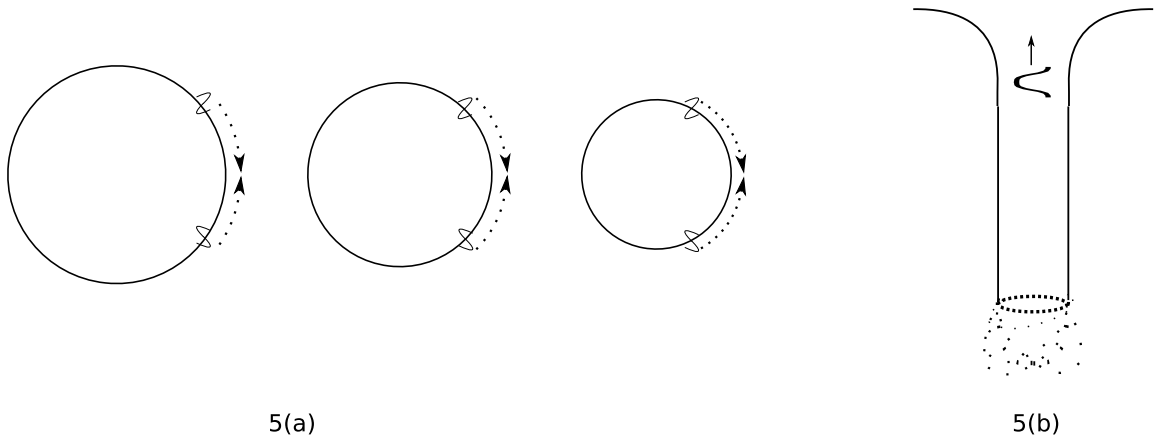} 
   
   \b

   \caption{ The state $\Psi_{out}$}
\end{figure}

  But our initial state $\Psi_{in}$ for the infalling particle does not evolve to $\Psi_{out}$.\footnote{The  relation between $\Psi_{in}, \Psi_{out}$ is reminiscent of `black hole complementarity' \cite{suss} where infalling and static observers describe states very differently.}  Instead, for the state $\Psi_{in}$ the 
collisions of excitations happen at different times on different loops. There is a phase cancellation in the emerging wave, and the excitation stays `trapped' on the loops for a long time \cite{gm}.  It can be easily checked (using the basic idea of fig.3) that when the excitations in $\Psi_{in}$ are separated by a distance of order the loop length (fig.4), in the gravity description  the quantum 
falls past the dashed circle of fig.2(d).\footnote{Recall that this circle was drawn at the point where the spread in loop lengths caused the throat to develop a complicated cap.}  Thus we can say that  the particle seems trapped behind the `horizon'.  

\b\b

\b\b

To summarize, we learn from field theory/gravity duality that `spacetime manifold' is only a coarse grained description of a complicated object, which for our case is described  in the dual field theory by a large number of loops. Semiclassical particle propagation requires that the excitations on each loop be synchronized to reflect the same position in the gravity dual. We can start with a state $\Psi_{in}$ of this type, but the highly entropic state of the hole `messes up' the synchronization between the loops, and we do not evolve to a state like $\Psi_{out}$ which is correctly synchronized to describe outgoing particles. There is no canonical choice of state in the horizon region, and thus no unique definition of spacetime manifold. 

While we have used the simplest black hole for our discussion (the 2-charge extremal hole), we expect that the fuzzball picture will hold for all holes,\footnote{See \cite{many} for some microstate constructions for other charges and dimensions, and for some studies of fuzzball properties.} so spacetime will  be `very quantum' and an ill-defined notion in the black hole interior.
 
\bigskip

\bigskip

{\bf Acknowledgements:}\quad I am grateful to Borun D. Chowdhury and Stefano Giusto for many helpful discussions. This work was supported in part by DOE grant DE-FG02-91ER-40690.


\newpage

\begin{thebibliography}{99}

\bibitem{maldacena}
  J.~M.~Maldacena,
  Adv.\ Theor.\ Math.\ Phys.\  {\bf 2}, 231 (1998)
  [Int.\ J.\ Theor.\ Phys.\  {\bf 38}, 1113 (1999)]
  [arXiv:hep-th/9711200].

\bibitem{maldasuss}
  J.~M.~Maldacena and L.~Susskind,
  Nucl.\ Phys.\  B {\bf 475}, 679 (1996)
  [arXiv:hep-th/9604042].
  
  \bibitem{lm4}
  O.~Lunin and S.~D.~Mathur,
  Nucl.\ Phys.\  B {\bf 623}, 342 (2002)
  [arXiv:hep-th/0109154].


\bibitem{sen}
  A.~Sen,
  Nucl.\ Phys.\  B {\bf 450}, 103 (1995)
  [arXiv:hep-th/9504027];
C.~Vafa,
  Nucl.\ Phys.\  B {\bf 463}, 435 (1996)
  [arXiv:hep-th/9512078].
  
  \bibitem{dabholkar}
  A.~Dabholkar,
  Phys.\ Rev.\ Lett.\  {\bf 94}, 241301 (2005)
  [arXiv:hep-th/0409148].

  
  
  \bibitem{lm5}
  O.~Lunin and S.~D.~Mathur,
  Phys.\ Rev.\ Lett.\  {\bf 88}, 211303 (2002)
  [arXiv:hep-th/0202072].
  
 \bibitem{suss}
  L.~Susskind, L.~Thorlacius and J.~Uglum,
  Phys.\ Rev.\  D {\bf 48}, 3743 (1993)
  [arXiv:hep-th/9306069];
  D.~A.~Lowe, J.~Polchinski, L.~Susskind, L.~Thorlacius and J.~Uglum,
  Phys.\ Rev.\  D {\bf 52}, 6997 (1995)
  [arXiv:hep-th/9506138].
  
  \bibitem{gm}
  S.~Giusto and S.~D.~Mathur,  {\it to appear}. 
  
 \bibitem{many}
 S.~Giusto, S.~D.~Mathur and A.~Saxena,
  Nucl.\ Phys.\  B {\bf 710}, 425 (2005)
  [arXiv:hep-th/0406103];
 I.~Bena and N.~P.~Warner,
  arXiv:hep-th/0701216;
 I.~Bena and N.~P.~Warner,
  Adv.\ Theor.\ Math.\ Phys.\  {\bf 9}, 667 (2005)
  [arXiv:hep-th/0408106];
  S.~D.~Mathur,
  Fortsch.\ Phys.\  {\bf 53}, 793 (2005)
  [arXiv:hep-th/0502050];
  V.~Balasubramanian, E.~G.~Gimon and T.~S.~Levi,
  arXiv:hep-th/0606118;
 I.~Kanitscheider, K.~Skenderis and M.~Taylor,
  arXiv:0704.0690 [hep-th].

  
\end{thebibliography}
\end{document}